\documentclass[aps,twocolumn]{revtex4}%
\usepackage{amsfonts}
\usepackage{amsmath}
\usepackage{amssymb}
\usepackage{graphicx}%
\begin{document}
\title[Classical Qubits]{A Classical Analysis of Capacitively Coupled Superconducting Qubits}
\author{James A. Blackburn$^{1}$, Jeffrey E. Marchese$^{2}$, Matteo Cirillo$^{3}$, and
Niels Gr{\o }nbech-Jensen$^{2}$}
\affiliation{$^{1)}$Department of Physics \& Computer Science, Wilfrid Laurier University,
Waterloo, Ontario N2L 3C5, Canada}
\affiliation{$^{2)}$Department of Applied Science, University of California, Davis,
California 95616}
\affiliation{$^{3)}$Dipartimento di Fisica and MINAS-Lab, Universit\`{a} di Roma "Tor
Vergata", I-00173 Roma, Italy}
\keywords{coupled qubits, Josephson}
\pacs{85.25.Cp, 74.50.+r, 03.67.Lx}

\begin{abstract}
An electrical circuit consisting of two capacitively coupled
inductive loops, each interrupted by a Josephson junction, is
analyzed through the classical RCSJ model. The same circuit has
recently been studied experimentally and the results were used to
demonstrate quantum mechanical entanglement in the system by
observing the correlated states of the two inductive loops after
initial microwave perturbations. Our classical analysis shows that
the observed phenomenon exists entirely within the classical RCSJ
model, and we provide a detailed intuitive description of the
transient dynamics responsible for the observations.

\end{abstract}
\maketitle

\bigskip
\section{Introduction}

In 2005, McDermott et al. \cite{McDermott} reported the experimental
observation of quantum mechanical entanglement in a system of two qubits, each
comprised of a superconducting loop interrupted by a single underdamped
Josephson junction, and weakly coupled via a capacitor. \ A second report with
some additional detail soon followed \cite{Steffen}, and a review by Siddiqi
and Clarke \cite{Clarke} of the key ideas appeared in that same issue of Science.

Briefly, the model put forward to explain the experimental observations was as
follows. The potential energy of a superconducting ring interrupted by a
Josephson junction, plotted as a function of the junction phase variable, can
possess a shallow well if biased with an appropriate magnetic field. Quantum
mechanics dictates discrete energy levels within a finite well, and a
sufficiently shallow well would have just a few permitted bound states, of
which the two lowest are denoted $\left\vert 0\right\rangle $, $\left\vert
1\right\rangle $. \ The argument has been advanced that such a system can be
viewed as a fictitious particle being limited to these quantum states
\cite{Clarke2}, rather than, as in a \textquotedblleft
classical\textquotedblright\ picture, freely exploring the potential surface.
This distinction is crucial to what follows.

In the experiments, two loops were subjected to a steady flux bias that made
the occupied wells shallow. One of the loops was given additional
time-dependent flux biasing in the form of a $10\operatorname{ns}$ microwave
burst. The microwave frequency was set to match the separation of the two
levels within the driven well, so it was imagined that the pulse would
selectively populate the first excited state of the driven loop, initially
putting the coupled system in the $\left\vert 01\right\rangle $ state. \ Then,
after a \textquotedblleft free evolution time\textquotedblright\ during which
the qubits interact, the states of both loops were probed using identical
measurement pulses of exactly the correct amplitude to reduce the barrier and
facilitate selective tunneling (or, more precisely, Macroscopic Quantum
Tunneling (MQT)) out of the upper state $\left\vert 1\right\rangle $, but not
out of the lower state $\left\vert 0\right\rangle $. Hence, if either well
were in an excited state, then escape via tunneling would lead to time
dependent phase dynamics and a resulting observed signal; if it were in the
lower state, no signal would be seen. This probing was repeated many times for
each selected $t_{free}$ and the accumulated data were argued to reveal the
probabilities for $\left\vert 00\right\rangle $, $\left\vert 01\right\rangle
$, $\left\vert 10\right\rangle $, and $\left\vert 11\right\rangle $, and to
confirm that entanglement had been observed.

We here provide a classical Resistively and Capacitively Shunted
Junction (RCSJ) analysis of the capacitively coupled qubit system. \
This builds upon our previous work which has reproduced several
experimental observations, such as those seen in \cite{Vion03},
including: multi-peaked distributions \cite{spectroscopy},
Rabi-oscillations \cite{Marchese}, Ramsey fringes \cite{Ramsey}, and
spin-echo \cite{spin-echo}. In this model the microwave pulse
stimulates small phase oscillations in the driven loop and, because
of the coupling, the passive loop soon develops its own phase
oscillations. \ Thus we study a straightforward picture -- that of a
pair of weakly coupled oscillators. \ Our analysis shows that these
coupled phase oscillations can yield very good quantitative
agreement with the core of experimental observations. It also points
to the crucial role of thermal noise in this system.

\section{The Model}
\begin{figure}
[pt]
\begin{center}
\includegraphics[
trim=0.000000in 0.103025in 0.000000in 0.068900in,
natheight=6.0000in,
natwidth=6.7510500in,
height=2.50000in,
width=2.510500in
]%
{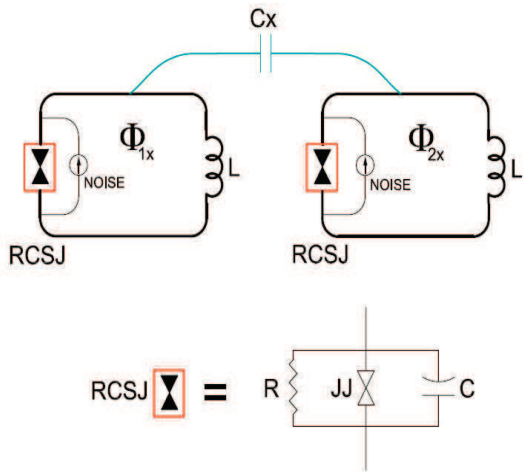}%
\caption{Configuration of two capacitively coupled superconducting loops, each
containing  a Josephson junction represented by the RCSJ equivalent
circuit and an associated noise source.}%
\label{Fig.1}%
\end{center}
\end{figure}
We begin with the circuit shown in Fig.~1. Each junction is
characterized by a critical current $I_{c}$, resistance $R$, and
capacitance $C$; each loop has an inductance $L$. \ The two loops
are coupled through a capacitance $C_{X}$. \ Let $\Phi_{1x}$ be the
externally applied flux on Loop \#1 and $\Phi_{2x}$ be the
externally applied flux on Loop \#2. \ With overdots denoting
derivatives in dimensionless time $\tau=\omega_{J}t$, and with the
junction plasma frequency $\omega_{J}=\sqrt{{2eI_{c}}/{\hbar C}}$,
the resulting equations of motion for the two junction phases can be
shown to be
\begin{eqnarray}
\ddot{\varphi}_{1}+\alpha\dot{\varphi}_{1}+\sin\varphi_{1} & = & \gamma_{x}%
(\ddot{\varphi}_{2}-\ddot{\varphi}_{1})-{\beta_{L}^{-1}}\left(  \varphi
_{1}+2\pi M_{1x}\right)  \nonumber \\ \label{Equation1} \\ %
\ddot{\varphi}_{2}+\alpha\dot{\varphi}_{2}+\sin\varphi_{2} & = & \gamma_{x}%
(\ddot{\varphi}_{1}-\ddot{\varphi}_{2})-{\beta_{L}^{-1}}\left(  \varphi
_{2}+2\pi M_{2x}\right)  \nonumber \\ \label{Equation2}%
\end{eqnarray}
where $\alpha=1/\omega_{J}CR$, and $\beta_{L}%
=2\pi{LI_{c}}/{\Phi_{0}}$; $\Phi_{0}$ being the flux quantum. $M_{ix}%
=\Phi_{ix}/{\Phi_{0}}$ is the normalized applied flux through loop $i$ and the
mutual coupling coefficient is $\gamma_{x}=C_{x}/C$. The characteristic energy
of this system is $E_{J}=I_{c}\Phi_{0}/2\pi$. The equations of motion can also
be put in the convenient form,
\begin{align}
\ddot{\varphi}_{a}+\alpha\dot{\varphi}_{a}+\sin\varphi_{a}\cos\varphi_{b}  &
=-{\beta_{L}^{-1}}\left[  \varphi_{a}+2\pi M_{a}\right] \label{Equation3}\\
g^{-1}\ddot{\varphi}_{b}+\alpha\dot{\varphi}_{b}+\sin\varphi_{b}\cos
\varphi_{a}  &  =-{\beta_{L}^{-1}}\left[  \varphi_{b}+2\pi M_{b}\right]  \;,
\label{Equation4}%
\end{align}
where $\varphi_{a}=(\varphi_{1}+\varphi_{2})/2$ and $\varphi_{b}=(\varphi
_{1}-\varphi_{2})/2$ are transformed variables, and $M_{a}=(M_{1x}%
+M_{2x})/2$, $M_{b}=(M_{1x}-M_{2x})/2$, and $g^{-1}={1+2\gamma_{x}}$ are the
corresponding magnetic fields and coupling, respectively. Based on published
data, we set the parameters at $\alpha=5\times10^{-5}$ (very light damping),
$\beta_{L}=2.841$, and $g=0.9954$, $I_{c}=1.1%
\operatorname{\mu A}%
$, $C=1.3\operatorname{pF}$,
$\omega_{J}^{-1}=0.02\operatorname{ns}$. Both loops are biased with
a dc flux $0.6941$ (resonance $\approx$5.1GHz) and with superimposed
pulses as shown in the upper panel of Fig.2.

\begin{figure}
[pt]
\begin{center}
\includegraphics[
trim=0.000000in 0.103025in 0.000000in 0.068900in,
natheight=3.250000in,
natwidth=2.510500in,
height=3.250000in,
width=2.510500in
]%
{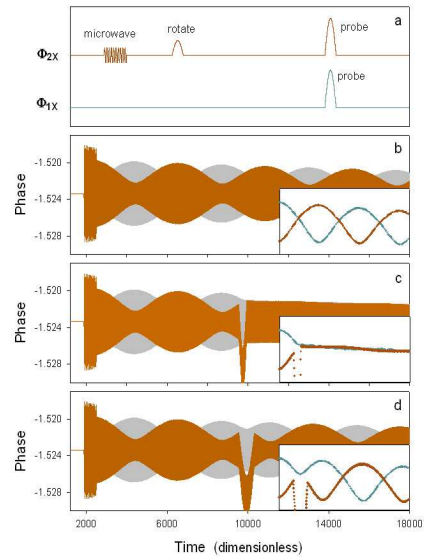}%
\caption{(a) Bias sequence for the two loops (not to scale). (b)-(d) Numerical
solutions of the coupled loop equations. \ In each plot, the darker
wave form is for Loop \#2, the lighter for Loop \#1. \ In (c) and
(d), a rotation pulse is applied at time 9550. \ The insets show
portions of the upper envelopes of the modulated phase oscillations.
}%
\label{Fig.2}%
\end{center}
\end{figure}

\section{Simulation Results}

We have numerically solved the coupled equations \ (\ref{Equation3}) and
(\ref{Equation4}) using both fourth order Runge-Kutta and Verlet algorithms.
Simulations at zero temperature were conducted as follows.%

The duration of the microwave burst was set at $100$ plasma periods;
the amplitude and normalized frequency were $0.000580$ and $0.989$,
respectively. Our results are shown in the three lower panels of
Fig.2. In each plot, the darker trace is the phase in Loop \#2. \
The effect of the microwave burst is easily seen -- the phase
$\varphi_{2}$ is kicked into rapid oscillations around its initial
rest value of $-1.52\operatorname{rad}$. These oscillations continue
after the microwave burst has ended. Because of the coupling, the
phase in Loop \#1 also oscillates. \ Furthermore, in a manner
characteristic of weakly coupled oscillators \cite{Chow}, each
waveform has the appearance of an amplitude modulated carrier, and
the envelope of one is exactly out of phase with the other. \ Note
that the amplitude of these modulations is very small compared to
the dc level. The junction phase oscillations are so rapid (the
frequency ratio of carrier to modulation is about $400\colon1$) that
on the scale of this figure they become compressed into solid
shading.

A rotation pulse of amplitude $0.000915$ was applied at time $9550$.
For these simulations, we chose widths of $0$, $60$, and $120$
plasma periods, equivalent to $0\operatorname{ns}$,
$7.5\operatorname{ns}$, and $15\operatorname{ns}$. The insets in
each of the three panels show the upper modulation envelopes as
functions of time and their responses to the different rotation
pulse widths. \ The significance of these plots is their exact
matching to the time dependence of the experimental probabilities
$P_{01}$ and $P_{10}$ given in Fig.~2B,C,D of \cite{Steffen}. \
Steffen et al. say: \textquotedblleft The occupation probabilities
$P_{01}$ and $P_{10}$ oscillate out of phase with a period of
$100\operatorname{ns}$, consistent with the spectroscopic
measurements.\textquotedblright\ Our numerical results exhibit the
same shapes and phase relationships for the envelopes. As well, our
simulation data show the period of the envelopes to be $87$ns. In
other words, the modulation envelopes that are a natural outcome of
the classical physics of weakly coupled oscillators are a match to
the experimental data, and the underlying dynamical phenomena of
both the modulation and the results of manipulation observed in
Ref.~\cite{Steffen} are well represented in the RCSJ model.

\begin{figure}
[pt]
\begin{center}
\includegraphics[
trim=0.000000in 0.103025in 0.000000in 0.068900in,
natheight=10.000400in,
natwidth=8.500200in,
height=4.000000in,
width=3.510500in
]
{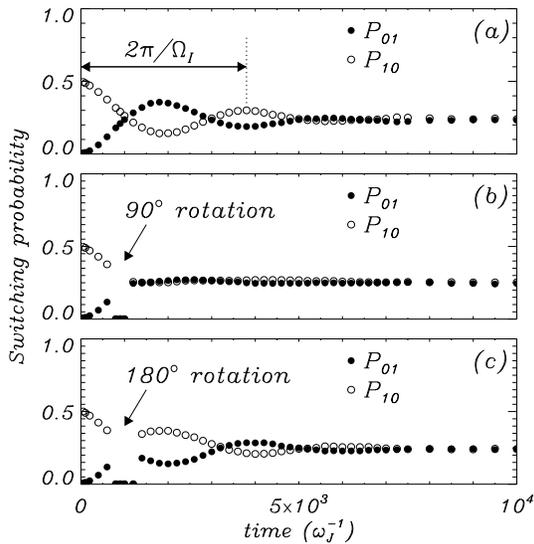}%
\caption{Numerical simulations for $T=25$ $mK$ with thermalization time of
10$^{3}$ time units, and $\alpha=$5$\times$10$^{-5}$. Pulsed
microwave frequency is 0.625, and pulse duration and amplitude are
507 and 0.00058, respectively. Each marker represents an average of
1,000 measurements conducted with a triangular probe pulse of
amplitude 0.0538 and width 500. (b) and (c) rotation pulse
application time is 818.75 after termination of the microwave, and
rotation amplitude is 0.000915. (b) 90$^{\circ}$ rotation pulse
duration is 253.5. (c) 180$^{\circ}$ rotation pulse duration is 507.}
\label{Fig.3}
\end{center}
\end{figure}

The experiments reported in \cite{Steffen} were conducted at $25$
$mK$. Following their experimental protocols, we added appropriate
noise to the system equations (\ref{Equation3}) and
(\ref{Equation4}), simulating $25mK$ thermal noise with a short
transient time of 1,000 time units. \ For any chosen moment
following the microwave burst ($t_{free}$), repeated simulation runs
were carried out and data were gathered on how often the individual
loops experienced an escape following the application of the probe
pulse (amplitude $0.0358$). The results are shown in Fig.3. The
close correspondence
with the reported experimental results in \cite{Steffen} is obvious.%

Comparing Fig.2 and Fig.3, it is apparent that in combination with
noise, the probe pulse effectively `teases out' the modulation
profiles of the junction phase oscillations. Equivalently, the
likelihood that the \textquotedblleft particle\textquotedblright\
will escape from its well will be higher at moments when the
noise-free oscillation amplitude (energy) is relatively large, and
conversely, when the oscillation amplitude is small, the chance of
escape is less. Hence the classical description of the system,
embodied in Eqs.(\ref{Equation3},\ref{Equation4}) captures all the
significant features observed in the experiments. However, there is
a wrinkle in the above picture that we will discuss after having
identified the dynamical modes in the system.

\section{Perturbation Analysis}

Let us first provide a simple analysis for the \textit{single} qubit.
Inserting the ansatz $\varphi=\varphi_{0}+A\sin{\omega t}$ into
Eq.~(\ref{Equation1}) for $\gamma_{x}=0$ we get the following relationships
between the fixed phase $\varphi_{0}$ and the resonance frequency $\omega_{r}$
to the magnetic field $M_{x}$ and amplitude $A$ of oscillation:
\begin{equation}
\omega_{r}^{2}=\beta_{L}^{-1}+(J_{0}(A)+J_{2}(A))\cos\varphi_{0}
\label{Equation5}%
\end{equation}
where
\begin{eqnarray}
J_{0}(A)\beta_{L}\sin\varphi_{0} & = & -\varphi_{0}-2\pi M_{x} \; . \nonumber
\end{eqnarray}
The latter expression provides the mean phase $\varphi_0$, while the former
Eq.(\ref{Equation5}) is the anharmonic
resonance shown in Fig.4a for small to moderate oscillation amplitudes.%

\begin{figure}
[pt]
\begin{center}
\includegraphics[
trim=0.000000in 0.647924in 0.000000in 1.947071in,
natheight=11.000400in,
natwidth=8.500200in,
height=3.2776in,
width=3.3114in
]
{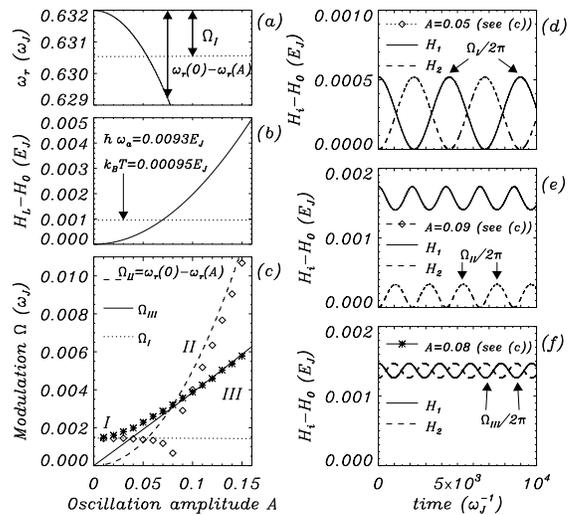}
\caption{Analyses of system modes with parameters given in the text. Single
loop resonance (a) and energy (b) as a function of oscillation
amplitude $A$. (c) Modulation frequencies; simulated (markers) and
perturbation analysis (lines). Different modes are engineered
through the initial conditions $\varphi_{i}(0)=\varphi_{0}+A_{i}$
such that mode-I is obtained by $A_{1}=A$ (small) and $A_{2}=0$,
mode-II by $A_{1}=A$ (large) and $A_{2}=0$, and mode-III by
$A_{1}\approx A_{2}$ (large). Examples of the three characteristic
modes observed in (c): mode-I (d), mode-II (e), and mode-III (f).}
\label{Fig.4}
\end{center}
\end{figure}

With this parameterized ansatz, we can provide the energy $H_{L}$ of a single
qubit as the averaged sum of kinetic energy and potential energy
$U(\varphi)=1-\cos\varphi+[\varphi+2\pi M_{x}]^{2}/2\beta_{L}$
\begin{align}
&  H_{L}=\langle\frac{1}{2}\dot{\varphi}^{2}+U(\varphi)\rangle=\frac{1}%
{4}A^{2}\omega^{2}+\label{Equation6}\\
&  \frac{1}{4\beta_{L}}A^{2}+1-J_{0}(A)\cos\varphi_{0}+\frac{1}{2\beta_{L}%
}[\varphi_{0}+2\pi M_{x}]^{2}\nonumber
\end{align}
This energy is shown in Fig.4b for the upper of the two potential
wells of the qubit potential. Notice that the energy of the
fixed-point is $H_{0}=U(\varphi_{0})$. With these basic anharmonic
relationships we can now proceed to analyzing the inherent
\textit{system} resonances.\\

\noindent\textbf{Mode-I: beat frequency mode.} In Eqs.~(\ref{Equation3}) and
(\ref{Equation4}), with $M_{b}=0$, $\alpha\approx0$, and small amplitude
oscillations of $\varphi_{1}$ and $\varphi_{2}$, we write $\varphi_{a}%
=\varphi_{0}+\psi_{a}$, where $\varphi_{0}$ is a constant and $|\psi_{a}|\ll
1$, and we will similarly assume $|\varphi_{b}|\ll1$. Inserting this ansatz
with $\alpha=0$ (for simplicity) gives the linear equations
\begin{align}
\ddot{\psi}_{a}+\sin\varphi_{0}+(\cos\varphi_{0}+\beta_{L}^{-1})\psi_{a}%
+\beta_{L}^{-1}\left(  \varphi_{0}+2\pi M_{a}\right)  =0  &  \label{Equation7}%
\\
g^{-1}\ddot{\varphi}_{b}+(\cos\varphi_{0}+\beta_{L}^{-1})\varphi_{b}=0\;.  &
\label{Equation8}%
\end{align}
From this we directly obtain $\beta_{L}\sin\varphi_{0}+\varphi_{0}+2\pi
M_{a}=0$, and we get the linear resonance frequencies $\omega_{a}$ and
$\omega_{b}$ of $\psi_{a}$ and $\varphi_{b}$, respectively, as $\omega_{a}%
^{2}=\cos\varphi_{0}+\beta_{L}^{-1}$ and
$\omega_{b}^{2}=g\omega_{a}^{2}$. Thus, in addition to the microwave
carrier frequency $\approx\omega_{a}$, we see a modulation frequency
at
\begin{eqnarray}
\Omega_{I} & = & \omega_{a}-\omega_{b}=(1-\sqrt {g})\omega_{a} \; ,
\label{Equation9}
\end{eqnarray}
which, with the reported experimental parameters, is given by
$\omega_{a}\approx0.63199$ and $\Omega_{I}\approx0.001455$. Notice
that $2\pi/\Omega_{I}$ is the exact modulation period (86.4ns) observed
in both experiments and simulations. The energy modulation of this
mode is exemplified in Fig.4d.\\

\noindent\textbf{Mode-II: phase-slip mode.} A direct consequence of
the anharmonicity of the system is that mode-I can only exist for
low system energy. The reason is that the two different oscillation
amplitudes ($A_{1},A_{2}$) observed in Fig.4d give rise to two
different resonance frequencies seen in Eq.~(\ref{Equation5}) and
Fig.4a. Thus, the linear mode-I cannot exist if the disparity
between the resonance frequencies overwhelms the linear modulation
frequency $\Omega_{I}$. If that happens, then the two oscillators
will shift to a mode-II, where the two loops will advance in two
different energy and frequency states, as illustrated in Fig.4e. The
observed energy modulation is a result of the mutual phase-slip
between the oscillators, which for systems of different oscillation
amplitudes $A_{1}>A_{2}$ can be approximated by
\begin{eqnarray}
\Omega_{II} & \approx & \omega _{r}(A_{2})-\omega_{r}(A_{1})
\; .
\label{Equation10}
\end{eqnarray}\\

\noindent\textbf{Mode-III: phase locked mode.} This is a mode,
illustrated in Fig.4f, which can be analyzed similarly to the
classical Rabi-type oscillation outlined in Ref.~\cite{Marchese}. We
define the ansatz{
$\varphi_{i}=\varphi_{0}(A_{i})+A_i\sin(\omega_{r}(A_{i})t+\theta_{i})$
($i=1,2$), $A_{1}\approx A_{2}\approx A$, and
$\theta_{1}=-\theta_{2}$. Considering the energy flow into qubit \#1
from the coupling,
\begin{eqnarray}
\dot{H}_{1} & = & \gamma_{x}
\dot{\varphi}_{1}(\ddot{\varphi}_{2}-\ddot{\varphi}_{1})
\label{Equation11} \\
& \approx & A\omega_r\cos(\omega_rt+\theta_1) \label{Eqatuion} \\
&& \times [A\omega_r^2\sin(\omega_rt+\theta_1)-A\omega_r^2\sin(\omega_rt-\theta_1)] \; ,
\nonumber
\end{eqnarray}
we write the energy
$H_{1}$ of qubit \#1 using Eq.~(\ref{Equation6}). The energy change $\Delta
H_{1}$ of qubit \#1 over one time unit can therefore be expressed
\begin{eqnarray}
&&\Delta H_{1}=\frac{\partial H_{1}}{\partial t}=\frac{\partial H_{1}}{\partial
A_{1}}\frac{\partial A}{\partial\omega_{r}}\ddot{\theta}_{1} \nonumber \\
&& =\langle\dot
{H}_{1}\rangle=\frac{1}{2}\gamma_{x}A_{1}^{2}\omega_{r}^{3}\sin(2\theta_{1}) \label{Equation13} \\
\Rightarrow && \frac{\partial H_{1}}{\partial
A_{1}}\frac{\partial A}{\partial\omega_{r}}\ddot{\theta}_{1}\approx\gamma_{x}A_{1}^{2}\omega_{r}^{3}\theta_{1} \label{Equation14} %
\end{eqnarray}
for $|\theta_{1}|\ll1$. This provides a slow modulation frequency
\begin{eqnarray}
\Omega_{III} & = & \sqrt{-\gamma_{x}A^{2}\omega^{3}
\frac{\partial\omega}{\partial A}/\frac{\partial H_{1}}{\partial A} }\; ,
\label{Eqation15}
\end{eqnarray}
where we note that $\partial\omega/\partial{A}<0$.\\

The above perturbation analysis provides definite expectations for
the type of excitations one can expect and how they relate to the
experimental observations. Judging from Figs.4d-f, we infer that
only mode-I can be the experimentally observed, due to the
statements in Ref.~\cite{Steffen} about using identical probe pulses
to get the presented high resolution and fidelity. The complexity of
the modulation resonances are shown in Fig.4c with comparisons
between evaluated modulation frequencies and direct numerical
simulations for $\alpha=0$ as a function of oscillation amplitudes.
We see that mode-I is always present for small $A$, and that it is
never present for large $A$ due the anharmonic modes II and III. The
transition to the anharmonic modes appears at a system energy given
by the the magnitude of $A$ -- see Figs.4c and 4b, where we have
also indicated the energy level of 25mK. It is here apparent that
mode-I does not exist with energy content much larger than the
thermal energy for the given system parameters. We have numerically
verified this for a fully thermalized system at 25mK. Only if not
fully thermalized (as is the case for Fig.3 above) or at
thermodynamic temperatures below 3mK do we observe mode-I.

\section{Discussion}

We note that results of the classical model depend on the choice of the
phenomenological damping parameter $\alpha$, which is not directly given by
the experimental data in \cite{Steffen}. The inverse of this parameter relates
to both the decay time for coherent signals in the system and the
characteristic time of thermalization; the value of this parameter is
therefore a significant component to understanding the system behavior. For
this presentation we have chosen a very small value of damping in line with
the observed decay times of the modulations as well as with the assumed large
sub-gap resistance associated with high quality aluminum junctions at the very
low temperature used for the experiments.

With the long history of the RCSJ model explaining experimental
observations, as outlined in the introduction, it is natural to
approach any new configuration from the same starting point. For the
present system, such direct classical analysis provides excellent
agreement with what is actually seen in the experiments, as
illustrated in Figs.2 and 3. However, our analysis also points to a
still unresolved question of thermal effects and the stability of
the relevant modes in the balance between temperature and
anharmonicity.

\begin{acknowledgments}
We are grateful for useful discussions with M.~R.~Samuelsen. This
work was supported in part (JAB) by a grant from the Natural
Sciences and Engineering Research Council of Canada.
\end{acknowledgments}

\end{document}